# Antiferromagnetic topological insulator with selectively gapped Dirac cones

A. Honma[1,#], D. Takane[1,#], S. Souma[2,3,*], K. Yamauchi[4], Y. Wang[5], K. Nakayama[1,6], K. Sugawara[1,3], M. Kitamura[7,8], K. Horiba[8], H. Kumigashira[9], K. Tanaka[10], T. K. Kim[11], C. Cacho[11], T. Oguchi[4], T. Takahashi[1,3], Yoichi Ando[5], and T. Sato[1,2,3,12,*]

[1]*Department of Physics, Graduate School of Science, Tohoku University, Sendai 980-8578, Japan*

[2]*Center for Science and Innovation in Spintronics (CSIS), Tohoku University, Sendai 980-8577, Japan*

[3]*Advanced Institute for Materials Research (WPI-AIMR), Tohoku University, Sendai 980-8577, Japan*

[4]*Center for Spintronics Research Network (CSRN), Osaka University, Toyonaka, Osaka 560–8531, Japan*

[5]*Institute of Physics II, University of Cologne, Köln 50937, Germany*

[6]*Precursory Research for Embryonic Science and Technology (PRESTO), Japan Science and Technology Agency (JST), Tokyo, 102-0076, Japan*

[7]*Institute of Materials Structure Science, High Energy Accelerator Research Organization (KEK), Tsukuba, Ibaraki 305-0801, Japan*

[8]*National Institutes for Quantum Science and Technology (QST),* Sendai 980-8579, Japan

[9]*Institute of Multidisciplinary Research for Advanced Materials (IMRAM), Tohoku University, Sendai 980-8577, Japan*

[10]*UVSOR Synchrotron Facility, Institute for Molecular Science, Okazaki 444-8585, Japan*

[11]*Diamond Light Source, Harwell Science and Innovation Campus, Didcot, Oxfordshire OX11 0QX, UK*

[12]*International Center for Synchrotron Radiation Innov1ation Smart (SRIS), Tohoku University, Sendai 980-8577, Japan*

[#]These authors contributed equally to this work.

*Corresponding authors.

**Abstract**

Antiferromagnetic (AF) topological materials offer a fertile ground to explore a variety of quantum phenomena such as axion magnetoelectric dynamics and chiral Majorana fermions. To realize such intriguing states, it is essential to establish a direct link between electronic states and topology in the AF phase, whereas this has been challenging because of the lack of a suitable materials platform. Here we report the experimental realization of the AF topological-insulator phase in NdBi. By using micro-focused angle-resolved photoemission spectroscopy, we discovered contrasting surface electronic states for two types of AF domains; the surface having the out-of-plane component in the AF-ordering vector displays Dirac-cone states with a gigantic energy gap, whereas the surface parallel to the AF-ordering vector hosts gapless Dirac states despite the time-reversal-symmetry breaking. The present results establish an essential role of combined symmetry to protect massless Dirac fermions under the presence of AF order and widen opportunities to realize exotic phenomena utilizing AF topological materials.

**Introduction**

To realize exotic quantum states in the topological insulator (TI), it is often necessary to break the time-reversal symmetry by introducing ferromagnetism into the crystal[1–4], as highlighted by the observation of quantum anomalous Hall effect[5–7]. As a natural extension, antiferromagnetic (AF) TIs are recently attracting particular attention because it is expected to show exotic properties[1,8–17], such as quantized magneto-electric effect accompanied by surface Hall conductivity[1,8,18,19], gigantic magneto-optical responses by AF fluctuations[9,20], and dynamic axion field useful for detecting dark-matter axions[21]. Also, AF TI is a useful platform applicable to spintronic devices owing to the ultrafast spin response and zero stray magnetic field[22,23]. To provide a pathway toward realizing exotic properties associated with antiferromagnetism and topology, it is essential to establish a new AF topological material, in particular, AF TI.

Prediction of AF TI was first made by Mong et al. in 2010 in their tight-binding model for the NaCl lattice with type-I anfiferromagnetism[2]. While the time-reversal symmetry ($\Theta$) is broken in the AF phase, the combined symmetry ($S = \Theta T_D$), where $T_D$ represents the translation by the **D** vector that inverts the spin direction (Fig. 1a), is preserved[8,12,13]. Materials having this $S$ symmetry are characterized by the $Z_2$ topological invariant as in the case of time-reversal-invariant TIs[8,12,13] and have been predicted to show a weak-TI-like behavior wherein the Dirac-cone SS is protected for the crystal plane parallel to the **D** vector, otherwise gapped. Thus, the AF TI has a unique experimental advantage distinct from strong TIs and ferromagnetic TIs; namely, the controllability of Dirac-cone SS through the manipulation of the **D** vector. This advantage manifests as intriguing characteristics of surface Dirac fermions; namely, the Dirac mass is strongly anisotropic and depends on the configuration of the AF structure, distinct from so-far established strong 3D TIs and topological semimetals. The protection by the $S$ symmetry in AF TI is a key ingredient to realize exotic properties[8–11,15,19,24]. It is thus essential to spectroscopically establish AF TI by distinguishing surfaces protecting/breaking $S$ symmetry.

As a candidate of AF TIs protected by $S$ symmetry, some materials such as MnBi$_{2n}$Te$_{3n+1}$ (MBT), EuIn$_2$As$_2$, and EuCd$_2$As$_2$ have been theoretically predicted[14,15,25–27]. The transport property that supports the topological nature of AF TIs such as the quantum anomalous Hall effect in MBT ($n$ = 1) has been intensively investigated[8,14,28], and various attempts to examine the associated magnetic gap of the Dirac-cone state in magnetic TI candidates have been made[14,29–33]. However, these materials have a layered structure and spectroscopies can access only a single surface parallel to the layer, making it difficult to obtain deep insights into the relationship between the AF order and topological SS. Also, in the case of MBT, there exists a fierce controversy on the magnitude and gapless/gapful nature of the Dirac-cone SS, which is not settled at the moment[34–40]. In this regard, the rare-earth monopnictide NdBi that exhibits type-I AF order (Fig. 1a) below Néel temperature of $T_N$ = 24 K is an excellent platform from the spectroscopic viewpoint, because (i) the cubic crystal structure is suited to access surfaces with different AF domains[41] and (ii) the large magnetic moment (3$\mu_B$) and large chemical ratio (50 %) of Nd ions may lead to a large magnetic Dirac gap. This provides us a precious opportunity to investigate the topological property unique to the AF TI, although the bulk semimetallic nature of monopnictides may hinder the quantum transport expected for AF TIs. A recent angle-resolved photoemission spectroscopy (ARPES) study on NdBi reported an unusual SS in the AF phase[42], making this material more interesting to study the entanglement between the AF order and electronic states (note that such SS was recently proposed to be a topologically trivial state and unlikely to be directly related to the AF TI properties[43]). Here, by utilizing the AF-domain-selective micro-focused ARPES, we established the electronic states of both $S$-preserving and $S$-broken surfaces with different Dirac-electron characteristics in NdBi.

**Results and Discussion**

$Z_2$ **topology in the paramagnetic phase.** First, we present the overall band structure of NdBi in the paramagnetic (PM) phase. Rare-earth monopnictide (RX$_p$) is a semimetal characterized by two hole pockets at the Γ point and electron pockets at the X point in the face-centered cubic (fcc) Brillouin zone (BZ) (Fig. 1b; refs.[44,45]). Bulk-sensitive ARPES with soft-X-ray photons together with first-principles band-structure calculations signifies bulk bands forming these pockets (Fig. 1c, d) together with the signature of bulk band inversion (for details, see Supplementary Note 1 and Supplementary Note 2). The Fermi-surface (FS) mapping in Fig. 1e obtained with vacuum ultraviolet (VUV) photons reveals corresponding features in the surface BZ. At the $\bar{\Gamma}$ point, we found inner square-like (h1) and outer diamond-like (h2) pockets which originate from the Bi-6$p$ hole bands (Fig. 1d). In Fig. 1e, one can also recognize vertically (e2) and horizontally (e1) elongated pockets at the $\bar{M}$ point originating from Nd-5$d$ bands at X points of different $k_z$'s ($k_z$ = 0 and 2$\pi$/$a$, respectively; Fig. 1b) due to the strong $k_z$ broadening of VUV photons[45–47]. The utilization of surface-sensitive VUV photons enables us to visualize the topological SS. One can identify in Fig. 1f a Dirac-cone band (called D1)[47–50] around the $\bar{\Gamma}$ point. While there is a single Dirac cone located at the binding energy $E_B$ of ~0.2 eV (which we call the Dirac-point energy $E_{DP}$) at the $\bar{\Gamma}$ point, the ARPES intensity across the $\bar{M}$ point in Fig. 1g signifies double Dirac-cone bands with $E_{DP}$ ~0.2 and ~0.4 eV (refs.[47–49,51]), called here D2 and D3, respectively. All these D1–D3 bands are confirmed to be of surface origin from their $h\nu$-independent energy position (for details, see Supplementary Note 3). The difference in the number of Dirac cones at two different time-reversal-invariant momenta (TRIM) at the surface corresponds to the difference in the number of band inversion in the bulk, namely, a single bulk X point is projected onto the $\bar{\Gamma}$ point whereas two inequivalent bulk X points onto the $\bar{M}$ point (see Fig. 1b). Odd numbers of Dirac cones in total suggest that NdBi in the PM phase is a $Z_2$ TI with a negative band gap (Fig. 1i), consistent with the parity analysis of the bulk-band structure which suggests the strong TI nature with $(\nu_0; \nu_1, \nu_2, \nu_3) = (1; 0, 0, 0)$; for details, see Supplementary Table S1. Our surface-projection calculations for the PM phase also reproduce the D1–D3 Dirac-cone states (Fig. 1h), consistent with the band picture obtained from the experiment (Fig. 1i; for details, see Supplementary Note 2).

**AF-induced reconstruction of the Dirac-cone SS.** Next, we show how the AF order influences the topological nature of NdBi. As summarized in Fig. 2a, the FS topology in the PM phase projected onto the surface BZ is characterized by the h1 and h2 hole pockets at $\bar{\Gamma}$, the D1 SS, the elongated e1 and e2 pockets at $\bar{M}$, and the D2 SS. As shown by the FS mapping around the $\bar{\Gamma}$ point at $T = 5$ K in Fig. 2b, all the pockets observed in the PM phase (h2, h1, and D1) are also resolved in the AF phase. A comparison of the band dispersion across the $\bar{\Gamma}$ point between the PM (Fig. 2c; $T = 30$ K) and AF phases (Fig. 2d; $T = 5$ K) shows that the D1 band is kept observed at both temperatures, indicating that the bulk-band inversion is preserved in the AF phase; however, a careful look at the D1 band reveals that the V-shaped upper branch (D1U) moves upward in the AF phase relative to that in the PM phase, while the $\Lambda$-shaped lower branch (D1L) moves downward, resulting in the opening of a Dirac gap associated with time-reversal-symmetry breaking due to the AF order. This gap is confirmed in more careful analyses of the energy distribution curve (EDC) at the $\bar{\Gamma}$ point (Fig. 2e) which signifies a double peak at $T = 5$ K as opposed to a single peak at $T = 30$ K, and its AF origin is also supported by the detailed temperature-dependent ARPES-intensity variation at the $\bar{\Gamma}$ point (Fig. 2f; for details, see Supplementary Note 4 and Supplementary Fig. 5). The magnitude of the Dirac gap at $\bar{\Gamma}$ estimated from the EDC is $125 \pm 5$ meV. This value is unexpectedly large despite the zero net magnetization and low $T_N$ (24 K) in NdBi and may be related to the effectively large local exchange field although the exact mechanism is unclear at the moment (for more detailed discussion, see Supplementary Note 4). Such a sizable Dirac gap that is directly linked to the AF order is in sharp contrast to that of an intrinsic magnetic TI candidate MBT where the Dirac gap is elusive[14,29–40].

Now we turn our attention to the AF-induced reconstruction of the other Dirac-cone bands located at the $\bar{M}$ point. Although the spectral features are complicated by the presence of two Dirac-cone bands D2 and D3 in the PM phase (Fig. 2h), these bands still survive in the AF phase (Fig 2i), as also seen in the FS mapping in the AF phase in Fig. 2g, where a small circular pocket originating from the upper branch of the D2 band (D2U) is observed inside the elongated bulk electron pockets, e1 and e2. The gapless X-shaped dispersion of the D2 band in the PM phase (Fig. 2h) turns into the upper (D2U) and lower (D2L) bands separated by a Dirac gap in the AF phase (Fig. 2i). This is also evident from the EDCs at the $\bar{M}$ point in Fig. 2j where a peak located at $E_B = 0.19$ eV at $T = 40$ K splits into two peaks at 0.12 and 0.28 eV at $T = 5$ K, exhibiting a Dirac gap of $160 \pm 5$ meV.

The Dirac gap opening for the D2 and D3 bands in the AF phase is also supported by our slab calculations for the AF phase (for details, see Supplementary Fig. 2).

To validate the AF origin of the observed spectral change, we have performed temperature-dependent ARPES measurements across $T_N$. The ARPES intensity at the $\bar{M}$ point plotted against temperature in Fig. 2k signifies no discernible change in the intensity profile above $T_N$, as evident from the $T$-invariant $E_B$ position of the D2 band. On the other hand, the D2 band splits into two bands as soon as the sample is cooled down below $T_N$. On lowering the temperature, the splitting is gradually enhanced and almost saturated below $T = 15$ K. This unambiguously demonstrates the AF origin of the Dirac gap. We note that the splitting of the D3 band was not clearly observed because of its weak intensity; the gap is likely much smaller than that of the D2 band (< 40 meV). This trend is also recognized in the slab calculation shown in Supplementary Fig. 2.

**Domain-selective electronic states in the AF phase.** Since NdBi crystal is expected to inherently contain multiple AF domains at the surface without magnetic field owing to the cubic structure (Fig. 1a), we surveyed the band structure in the AF phase by scanning a micro-focused beam on the surface and found that there exists another domain (called domain B) that exhibits a spectral feature markedly different from that discussed above (called domain A). As shown in Fig. 3a, on domain A, the bulk h1 and h2 bands outside the D1 band smoothly disperse toward $E_F$ without anomalies, resembling the band dispersion of the PM phase (Fig. 1f). On the other hand, the energy bands of domain B look significantly reconstructed (Fig. 3b). While the h1 band shows a similar dispersion for two domains, the broad feature originating from the h2 band seen in domain A (Fig. 3a) turns into a couple of sharp features (S1 and S2; white arrows) crossing $E_F$ with a shallower dispersion. The S1 and S2 bands were assigned to the magnetically split Fermi-arc SS[42].

The definitive domain-dependent nature also shows up in the Dirac-cone SS, D1. Although domain A hosts a Dirac gap of 125 meV (Figs. 3a and 2d), an X-shaped band with no signature for a Dirac gap is observed in domain B (Fig. 3b). A critical difference is also visualized by a comparison of the EDC at $T = 5$ K in Fig. 3c which signifies a double peak for domain A in contrast to a single peak for domain B. This suggests symmetry protection of the Dirac-cone SS (D1) in domain B despite the time-reversal-symmetry breaking. In contrast to the D1 band, the D2 band at the $\bar{M}$ point gapped out by 160 meV for domain A (Fig. 3d) still shows the band separation (D2U and D2L) by 80 meV for domain B (Fig. 3e). Persistence of the gap of the D2 band is also visualized by the EDCs (Fig. 3f) which signify the double-peaked structure (D2U and D2L) for both domains (note that the D3 peak at $E_B \sim 0.4$ eV is sharper for domain B than domain A, indicative of a change in the gap magnitude, although its quantitative estimation is difficult). According to the $Z_2$ classification, when the number of Dirac-cone SS is odd on a given surface, the Dirac cone is topologically protected, whereas it is not protected for the even-number case[8,12,13]. In the present case for domain B, the single Dirac-cone SS (D1) is protected, whereas the other two Dirac-cone SSs (D2 and D3) are not. Thus, the residual gap opening for the D2 band may not violate the $Z_2$ topological protection, but indicates its topologically more fragile nature. This point needs to be further examined, as detailed in Supplementary Note 2.

We systematically scanned micro-focused VUV photons on the cleaved surface (Fig. 3g), and probed the local band structure as a function of real-space position ($x$, $y$) in the AF phase, as highlighted by the mapping of photoelectron intensity around the $\bar{\Gamma}$ point (Fig. 3h). As a result, the ARPES spectra in the AF phase were categorized into either domain A, domain B, or other indistinguishable region, as indicated by different colorings in Fig. 3i. The distinction of domains was made by looking at the local band dispersion together with specifying the aforementioned anomalies in the h2 and D1 bands, as exemplified in Fig. 3j. We found that the ARPES spectrum in the PM phase shows no meaningful ($x$, $y$) dependence, indicative of the disappearance of such domains, confirming the AF origin of domains A and B. Such two types of domains have not been resolved in the previous ARPES study of NdBi[42], while some micro-ARPES studies have been already applied to other rare-earth pnictides such as CeSb[52,53]. The existence of multiple AF domains in NdBi was also confirmed by our polarizing microscope measurements, as detailed in Supplementary Note 5 and Supplementary Fig. 6.

**_S_-symmetry protection of the Dirac-cone SS.** To obtain further insights into the origin of intriguing domain-dependent electronic structure, we directly compare the FS topology between domains A and B as shown in Fig. 4a–d. By a side-by-side comparison of the FS around the $\bar{M}$ point (Fig. 4b, d), one can recognize that there exist small pockets at both corners of the horizontally elongated electron pocket, e1, only for domain B (white arrows). In fact, the ARPES intensity along a **k** cut crossing this pocket (red lines in Fig. 4b, d) signifies an additional shallow electron band that likely arises from the SS (S4)[42,43] besides the e1 band (inset to Fig. 4d) whereas it is completely absent for domain A (inset to Fig. 4b). Around the $\bar{\Gamma}$ point, while domain A shows a normal FS image (Fig. 4a) similar to the case of PM phase (Fig. 1e), domain B obviously has an anomaly at both corners of the h2 pocket (Fig. 4c); a small pocket (S2) that resembles the pocket of the $\bar{M}$-centered FS (Fig. 4d) seems to appear.

The observed critical differences in the FS topology between the two domains are explained by taking into account the existence of two types of AF domains at the same surface and the relevant band reconstruction, as highlighted in Fig. 4e–h. Our data for domain A are consistent with the magnetic structure in which the AF stacking occurs along the out-of-plane direction and the magnetic moment of Nd ions at the topmost surface layer aligns ferromagnetically (Fig. 4g). Since the observed bulk-band structure is significantly broadened along $k_z$ already in the PM phase, it does not show an effective change even when entering into the AF phase because the band folding occurs along the $k_z$ direction (Fig. 4e). The $C_4$ symmetric pattern in Fig. 4a also supports the AF-ordering along $k_z$. On the other hand, our data for domain B are consistent with the magnetic structure in which the AF stacking occurs along the in-plane direction (e.g. along the $x$-axis), corresponding to the antiparallel configuration of the magnetic moment at the topmost layer (Fig. 4h). In this case, the band folding occurs with respect to the magnetic BZ boundary at halfway between the $\bar{\Gamma}$ and $\bar{M}$ points ($k_x = \pi/a$; Fig. 4f). Consequently, the e1 pocket around $\bar{M}$ is folded to $\bar{\Gamma}$, and the h1 and h2 pockets around $\bar{\Gamma}$ to $\bar{M}$ (note that the intensity of the folded bulk bands is weak). Because the magnetic BZ has the $C_2$ symmetry along the $k_x$ axis, small pockets associated with the AF-induced SS (S1–S4) appear only at the horizontal side of the original bulk pocket, reflecting the magnetic BZ boundary (see Supplementary Note 6 for details). We have confirmed that the $C_2$ symmetric electronic structure is not associated with the matrix-element effect of photoelectron intensity, as detailed in Supplementary Note 7. Our argument is further corroborated by the domain-selective ARPES measurements on a cousin material NdSb[54,55] which signified the existence of all three types of AF domains (out-of-plane, in-plane horizontal/vertical) together with the non-topological (non $Z_2$) nature of the S1–S4 pockets. This conclusion is different from that of the recent study which suggests the $C_4$ symmetric FS[42], probably because of the better spatial resolution of the present study.

The configuration of the AF order plays a crucial role in the massive vs massless characteristics of the Dirac-cone SS. When the translation vector (**D**) lies on the surface (domain B; Fig. 4h), the combined symmetry $S = \Theta T_D$ is expected to be preserved and the Dirac cone is protected[8,12,13]. As we demonstrated in Fig. 3a–f, although the time-reversal symmetry ($\Theta$) is broken in the AF phase, the D1 band for domain B is still protected by the $S$ symmetry and maintains the massless character (Fig. 4j), in contrast to the massive character for domain A whose surface breaks the $S$ symmetry (Fig. 4i). Such distinction, which owes to the high spatial resolution of micro-focused ARPES, together with our first-principles band-structure calculations that signify the domain-dependent Dirac gap (for details, see Supplementary Note 2), firmly verifies the $S$-symmetry protection of the Dirac cone, and thus validates the long-awaited AF TI phase proposed by theory[8,9].

Topological protection by the $S$ symmetry suggests a high tunability of the Dirac-cone SS via controlling the AF domain and the surface index[8,15,19]. Such tunability would be useful for realizing exotic topological matters such as 3D axion insulators and high-order TI. For example, when we cut a NdBi crystal so as to make all facets have a domain-A-type configuration where the **D** vector contains a finite out-of-plane component, an axion insulator with a negative band gap may be realized[8,15,19]. Further, on a certain crystal hinge between these facets, a 1D chiral hinge state may emerge within the surface Dirac gap — a signature of higher-order TI[11,15,19]. At the step edge of the ferromagnetic layer, 1D chiral edge mode is proposed to appear[8,20,25]. Thus, besides the significance of the discovery of AF TI protected by the $S$ symmetry, NdBi also serves as a useful platform to realize exotic quantum states.

**Methods**

**Sample fabrication.** Single crystals of NdBi were grown by the flux method using indium flux. The raw materials were mixed in a molar ratio of Nd: Bi: In = 1: 1: 20 and placed in an alumina crucible. The crucible was sealed in an evacuated Quartz tube filled with Ar gas of 50 mbar. The ampoule was heated to 1100 °C, kept for 10 hours, and then cooled to 700 °C in 160 h. The excessive indium was removed in a centrifuge. Obtained crystals were characterized by X-ray diffraction measurements.

**ARPES measurements**. SX-ARPES measurements were performed with an Omicron-Scienta SES2002 electron analyzer with energy-tunable synchrotron light at BL2 in Photon Factory (PF), KEK. We used linearly polarized light (horizontal polarization) of 515–601 eV. VUV-ARPES measurements were performed with micro-focused VUV synchrotron light at BL28 in PF[56], BL5U in UVSOR, and I05 in Diamond Light Source. We used linearly or circularly polarized light of 60–75 eV. The energy resolution for the SX- and VUV-ARPES measurements was set to be 150 and 10–20 meV, respectively. Samples were cleaved in situ along the (001) plane of the cubic crystal in an ultrahigh vacuum of $1\times10^{-10}$ Torr. Prior to the ARPES measurement, the crystal orientation was determined by the X-ray Laue backscattering measurement which signifies clear four-fold symmetric diffraction spots consistent with the (001) cleaved plane. The Fermi level ($E_F$) of samples was referenced to that of a gold film electrically in contact with the sample holder.

**Calculations.** First-principles band-structure calculations were carried out by using a projector augmented wave method implemented in Vienna Abinitio Simulation Package (VASP) code[57]. To calculate the band structure for the PM phase, the modified Becke-Johnson (mBJ) potential[58] which is known to properly reproduce the band gap in $RX_p$[59], was used for the exchange-correlation functional. The total energy was calculated self-consistently with the tetrahedron sampling of 8×8×1 $k$-point mesh taking into account SOC. The surface states were obtained with the surface Green's function method implemented in WannierTools code[60] after the maximally localized Wannier functions for Bi-$s$, Bi-$p$, and Nd-$d$ orbital states were obtained by using Wannier90 code[61]. For the AF phase, we have carried out slab calculations with 12 atomic-layer slabs by taking into account the actual type-I AF structure (Supplementary Fig. 2 and 3). To properly take into account the magnetic moment of Nd ions, we included the strong correlation effect of Nd 4$f$ electrons by using GGA+$U$ calculation potential[62] instead of mBJ potential which has a nonconvergence problem in the slab calculations as in the previous study[57]. We have carried out the unfolding of bands for the superstructure in the AF phase for domain B by using a method proposed in the previous literature[63].

**Polarizing microscopy**. Polarizing microscopy measurements were performed by using a home-built UHV microscope system at Tohoku University. We have used a 100 W halogen lamp (U-LH100L-3, Olympus) to obtain bright reflectance images. Polarizing images were obtained in the crossed Nicols configuration with the optical principal axes along [110] axis. Samples were cleaved in situ along the (001) plane of cubic crystal in a UHV of $1\times10^{-10}$ Torr. The sample was cooled by liquid helium cryostat and the temperature was controlled in the range of 5–30 K.

**Data availability**

The data that support the findings of this study are available within the main text and Supplementary Information. Any other relevant data are available from the corresponding authors upon request.

**Acknowledgments**

We thank Y. Kubota, T. Kato, T. Kawakami, and N. Watanabe for their assistance in the ARPES experiments. This work was supported by JST-CREST (No. JPMJCR18T1), JST-PRESTO (No. JPMJPR18L7), Grant-in-Aid for Scientific Research (JSPS KAKENHI Grant Numbers JP21H04435 and JP19H01845), Grant-in-Aid for JSPS Research Fellow (No: JP23KJ0210 and JP18J20058), KEK-PF (Proposal number: 2021S2-001 and 2022G652), and UVSOR (Proposal number: 21-658 and 21-847). The work in Cologne was funded by the Deutsche Forschungsgemeinschaft (DFG, German Research Foundation) - Project number 277146847 - CRC 1238 (Subproject A04). A.H. thanks GP-Spin and JSPS, and D.T. thanks JSPS and Tohoku University Division for Interdisciplinary Advanced Research and Education.

**Author Contributions Statement**

The work was planned and proceeded by discussion among A.H., D.T., S.S., Y.A. and T.S. A.H., D.T., S.S., K.N., M.K., K.H., H.K., K.T., T.K., C.C., T.T and T.S. performed the ARPES measurements. D.T., Y.W., and Y.A. carried out the crystal growth. K.Y. and T.O. carried out first-principles band calculations. A.H., K.S. and S.S. performed the polarizing microscopy experiments. A.H., S.S., and T.S. finalized the manuscript with inputs from all the authors.

**Competing Interests Statement**

The authors declare no competing interests.

**ADDITIONAL INFORMATION**

**Supplementary information** is available for this paper at http://...

**Correspondence** and requests for materials should be addressed to S. S. (s.souma@arpes.phys.tohoku.ac.jp) and T. S. (e-mail: t-sato@arpes.phys.tohoku.ac.jp).

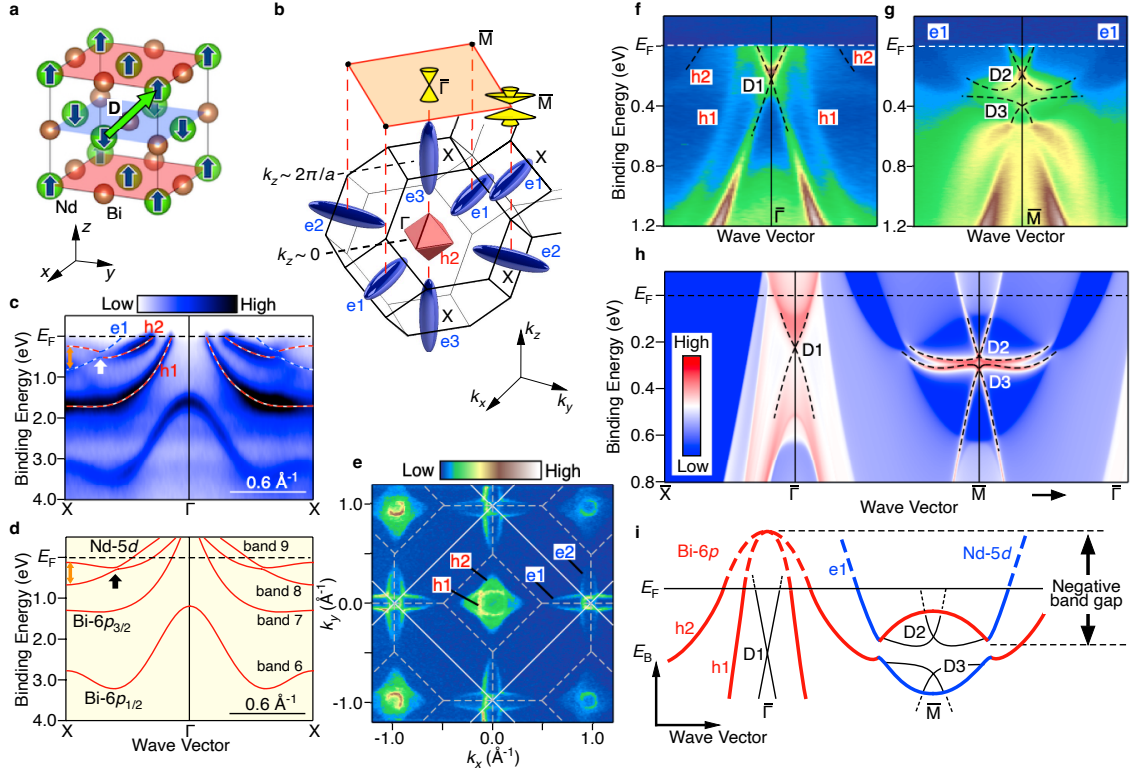

**Fig. 1| Band structure of NdBi in the paramagnetic phase. a** Rock-salt crystal structure of NdBi with spin orientation in the antiferromagnetic (AF) phase. Translation **D** vector is also indicated by the green arrow. **b** Schematic Fermi surface (FS) and bulk fcc Brillouin zone (BZ) of NdBi, together with the surface BZ projected onto the (001) plane (orange rectangle) and Dirac-cone surface state (SS) at the $\bar{\Gamma}$ and $\bar{M}$ points. **c** Angle-resolved photoemission spectroscopy (ARPES) intensity at $T = 40$ K measured along the $\Gamma X$ cut ($k_z \sim 0$) of bulk BZ with soft X-ray photons of $h\nu = 515$ eV. Red and blue dashed curves highlight the band dispersion for the Bi-6$p$ (h1, h2) and Nd-5$d$ (e1) orbital. The spin-orbit gap at the intersection of these bands and the band inversion at the X point are indicated by white and orange arrows, respectively. **d** Corresponding calculated bulk band structure in the paramagnetic (PM) state. The label of bands (6–9) is also indicated. **e** ARPES-intensity mapping at $E_F$ as a function of $k_x$ and $k_y$ at $T = 35$ K measured at $h\nu = 60$ eV. **f** ARPES intensity around the $\bar{\Gamma}$ point of surface BZ measured at $h\nu = 75$ eV. **g** Same as **f** but measured around the $\bar{M}$ point at $h\nu = 60$ eV. Black dashed curves in **f** and **g** are guides for the eyes to trace the surface band dispersions. **h** Calculated surface spectral weight along the $\bar{\Gamma}\bar{X}$ and $\bar{\Gamma}\bar{M}$ cuts projected onto the (001) plane, obtained with the Green-function method for a semi-infinite slab of NdBi in the PM phase. Black dashed curves trace the band dispersion of D1–D3 SS. **i** Schematic band structure of NdBi in the PM phase. D1, D2, and D3 indicated by thin black curves represent topological SS, whereas red and blue curves are bulk Bi-6$p$ bands forming hole pockets (h1 and h2) at $\bar{\Gamma}$ and Nd-5$d$ band forming an electron pocket (e1) at $\bar{M}$, respectively.

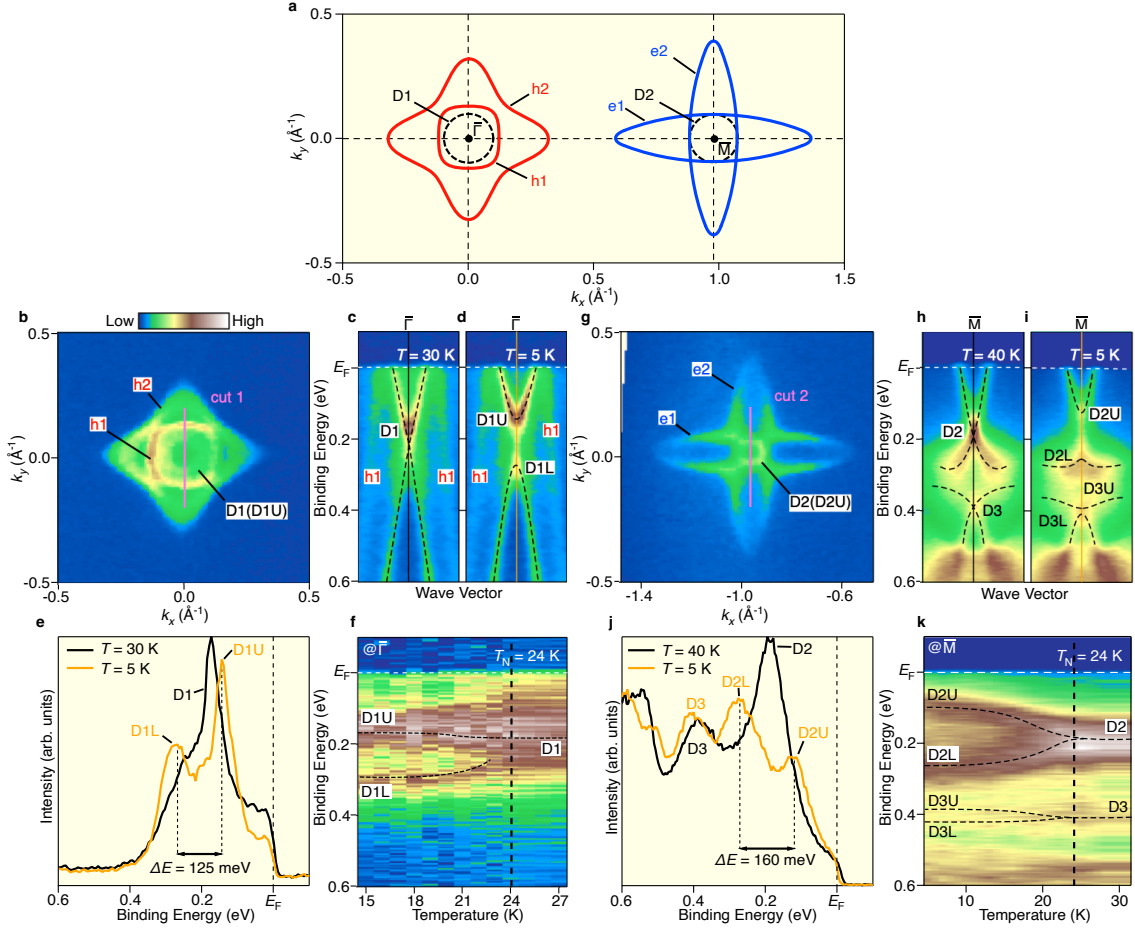

**Fig. 2| Observation of massive Dirac-cone bands in the AF phase. a** Schematic FS of NdBi projected onto the surface BZ. e1 and e2 (blue curves) represent electron pockets elongated along $k_x$ and $k_y$, respectively, and are located at different X points in the bulk BZ. h1 and h2 (red curves) represent inner and outer hole pockets, respectively. **b** FS mapping around the $\bar{\Gamma}$ point in the AF phase ($T = 5$ K) measured at $h\nu = 60$ eV. **c, d** ARPES intensity around the $\bar{\Gamma}$ point in the PM phase ($T = 30$ K) and the AF phase ($T = 5$ K), respectively, measured at $h\nu = 75$ eV. **e** EDCs at the $\bar{\Gamma}$ point at $T = 5$ K (orange curve) and 30 K (black curve). Dashed lines are a guide for the eyes to trace the upper and lower D1 (D1U and D1L). **f** Temperature dependence of ARPES intensity at the $\bar{\Gamma}$ point. Dashed curves are a guide for the eyes to highlight the gap opening across $T_N$ for D1. **g–i** Same as **b–d** but around the $\bar{M}$ point obtained at $h\nu = 60$ eV. Dashed curves in **h** and **i** are a guide for the eyes to trace the band dispersion of D2 and D3 as well as their lower and upper branches. **j, l** Same as **e** and **f** but measured at the $\bar{M}$ point. Dashed curves in **k** are a guide for the eyes to highlight the gap opening across $T_N$ for D2 and D3.

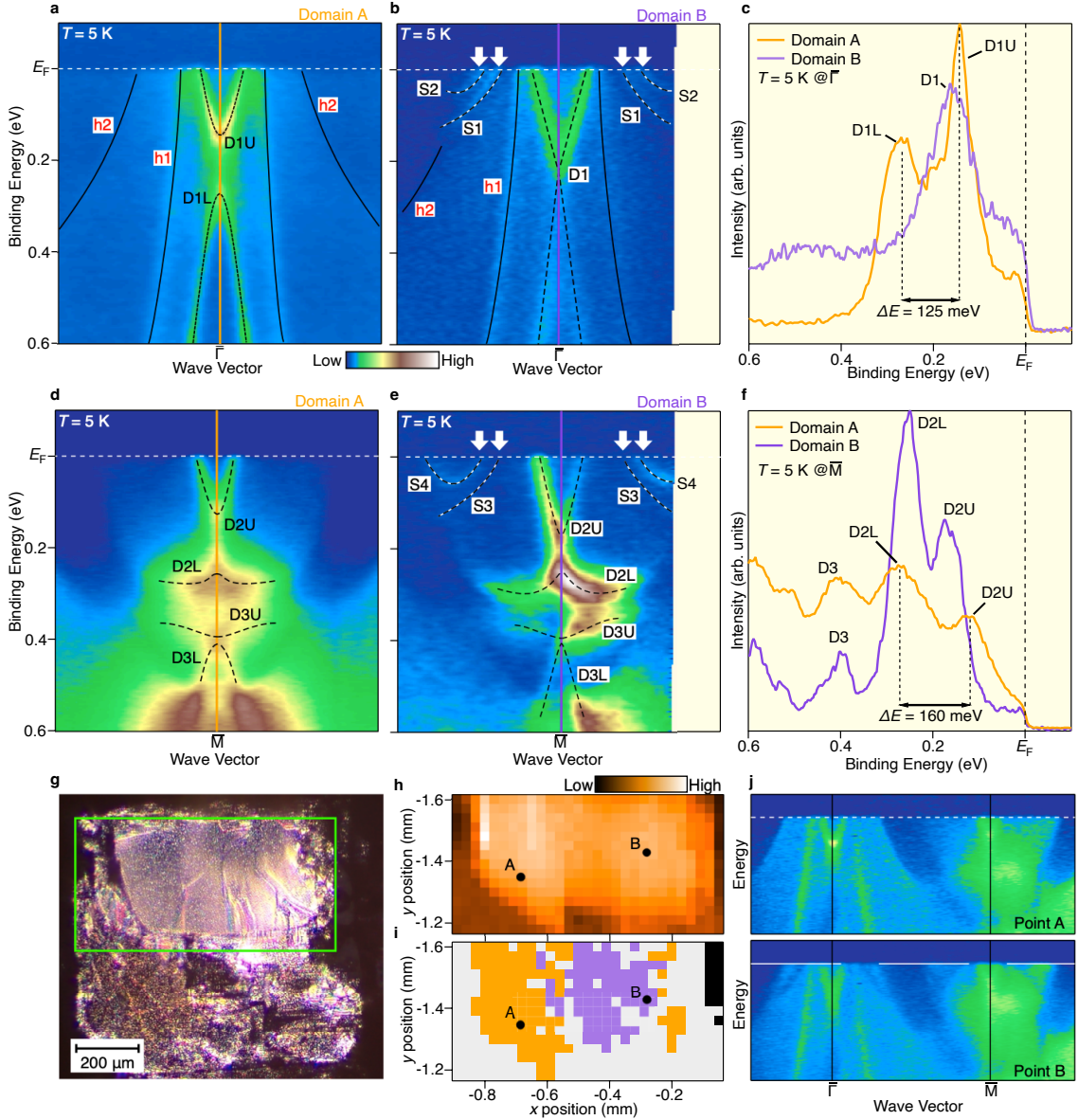

**Fig. 3| Domain-dependent band structure of NdBi. a**, **b** ARPES intensity around the $\bar{\Gamma}$ point for domains A and B, respectively. **c** Comparison of EDC at the $\bar{\Gamma}$ point at $T = 5$ K between two domains (orange for domain A and purple for domain B). **d**–**f** Same as **a**–**c**, but at the $\bar{M}$ point. **g** Optical microscope image for a cleaved surface of NdBi where domain-selective micro-ARPES measurements were performed. **h** Spatial mapping of the ARPES intensity integrated the ($E$, **k**) area around the $\bar{\Gamma}$ point, measured for the area enclosed by the green rectangle in **g**. **i** Distribution of domains A (orange) and B (purple) estimated from the spatially resolved ARPES-derived band structure. Gray and black colors represent indistinguishable and very-weak-intensity areas, respectively. **j** Representative ARPES intensity in the AF state obtained at specific real-space points A and B in **h** and **i**. Note that the statistics in **j** are poorer because many lateral positions with a mesh of 34 × 18 (in total ~600 data points) had to be covered for the spatial mapping.

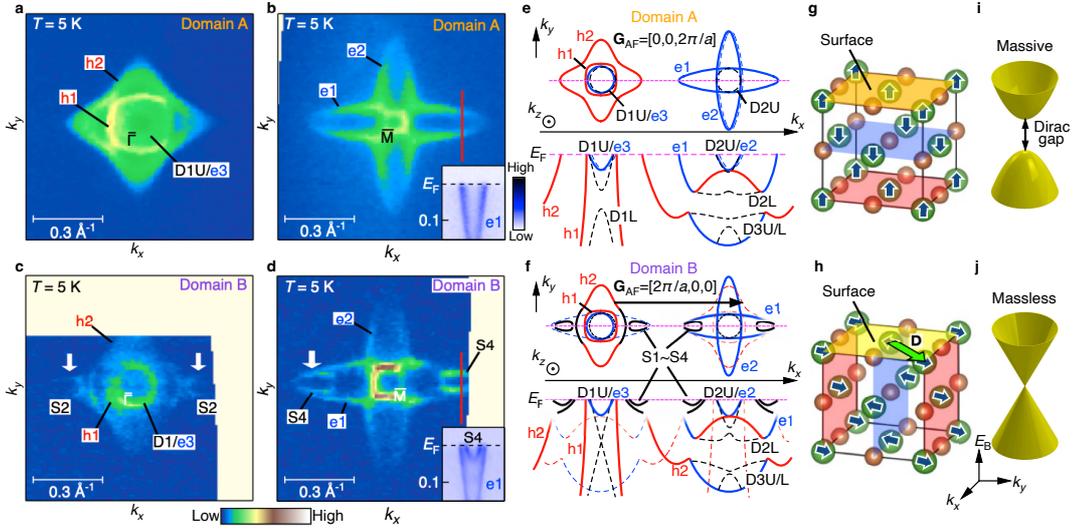

**Fig. 4| Schematics of the relationship between two AF domains and electronic states.**
**a**, **b** ARPES intensity around the $\bar{\Gamma}$ and $\bar{M}$ points, respectively, for domain A. Inset to **b** shows the near-$E_F$ ARPES intensity along a **k** cut shown by the red line. **c**, **d** Same as **a** and **b** but for domain B. **e**, **f** Schematics of the influence of AF structure to the folding of FS and band dispersion for domains A and B, respectively. $\mathbf{G}_{AF}$ is a reciprocal lattice vector of magnetic BZ for each domain. Red, blue, and black curves represent energy bands for bulk hole, bulk electron, and surface bands, respectively. **g**, **h** AF structure for domains A and B, respectively. Translation vector **D** inverting the spin direction is indicated by the green arrow in **h**. **i**, **j** Corresponding schematic band dispersion of Dirac-cone bands for domains A and B, respectively.

SUPPLEMENTARY INFORMATION for

"Antiferromagnetic topological insulator with selectively gapped Dirac cones" by A. Honma et al.

**Supplementary Note 1**: **Bulk electronic structure and band inversion of NdBi**

To visualize the bulk bands of NdBi in three-dimensional (3D) $k$ space, it is useful to use bulk-sensitive soft-X-ray (SX) photons because the longer photoelectron mean-free path relative to that for vacuum ultraviolet (VUV) photons reduces the intrinsic uncertainty of the out-of-plane wave vector $k_z$ through the Heisenberg's uncertainty principle and as a result allows the accurate 3D band mapping. Supplementary Fig. 1a and 1b show ARPES-intensity mapping at $E_F$ as a function of in-plane wave vector measured at $T = 40$ K [paramagnetic (PM) phase] for two representative $k_z$ slices in the bulk Brillouin zone (BZ) (see Fig. 1b of the main text) at $k_z \sim 2\pi/a$ ($h\nu = 601$ eV) and $\sim 0$ ($h\nu = 515$ eV), respectively. One can identify different intensity distributions between $k_z \sim 0$ and $2\pi/a$, signifying that the ARPES signal actually reflects the bulk Fermi surface. At $k_z \sim 0$ (Supplementary Fig. 1a), we find bright intensity spots centered at the Γ point associated with the bulk inner (h1) and outer (h2) hole pockets, together with a weaker intensity centered at the X point elongated along the ΓX direction attributable to the bulk electron pockets (e1 and e2). The hole and electron pockets originate from the topmost bulk valence bands with the Bi-6$p$ orbital character and the lowest bulk conduction band with the Nd-5$d$ character, respectively, which are responsible for the semimetallic nature of rare-earth monopnictides[1] as schematically shown in Fig. 1i of the main text. Since the bulk-band inversion at the X point is known to be directly linked to the topological nature[2–9], we show in Supplementary Fig. 1c (left panel) the ARPES intensity and (right panel) the corresponding second derivative intensity along the ΓX cut of bulk 3D BZ measured at $h\nu = 515$ eV. The result signifies a weak Nd 5$d$ electron band e1 (dashed blue curve) which crosses $E_F$ midway between the Γ and X points, together with the inner (h1) and outer (h2) Bi-6$p$ hole bands around the Γ point. The e1 and h2 bands appear to show a hybridization gap at the intersecting point (white arrow), suggestive of the inverted band structure as in the case of LaBi and CeBi[4,9].

To signify the correspondence between the bulk band inversion and the Dirac-cone surface state (SS), we have mapped the ARPES intensity with surface-sensitive VUV photons ($h\nu = 60$ eV) along the same ΓX cut (corresponding to the $\overline{\Gamma}\overline{M}$ cut in the surface BZ). Although the intensity of h1 and h2 bands is broadly distributed due to the short photoelectron escape depth and resultant strong $k_z$ broadening[2,9,10], one can clearly identify in Supplementary Fig. 1d the band-inversion-originated anomaly at the intersection of the Nd 5$d$ and Bi-6$p$ bands (white arrows) which reflects the hybridization of the e1 and h2 bands. Owing to the surface sensitivity of VUV photons, one can identify a weak Dirac-cone band (D1) around the $\overline{\Gamma}$ point inside the h1 band, as well as two Dirac-cone bands (D2 and D3) around the $\overline{M}$ point. Those bands (D1–D3) are not well resolved in the bulk-sensitive SX data (Supplementary Fig. 1c), consistent with their surface origin. Energy levels of Nd 5$d$ and Bi-6$p$ bands are inverted at the X point, one of the time-reversal-invariant momentum (TRIM), producing the Dirac-cone SS protected by time-reversal symmetry[2–6,9]. Associated with the difference in the number of bulk-band inversions projected onto the high-symmetry **k** points in the surface BZ, the number of Dirac-cone SS are different between the $\overline{\Gamma}$ and $\overline{M}$ points. Specifically, the D1 band at $\overline{\Gamma}$ is associated with the band inversion at single X point [**k** = (0, 0, $2\pi/a$)] of the bulk BZ, whereas the D2 and D3 bands are with the band inversions at two X points [**k** = ($2\pi/a$, 0, 0) and ($2\pi/a$, 0, $2\pi/a$)], as also explained in the main text (Fig. 1b). The odd number of total Dirac-cone SSs at TRIM, $\overline{\Gamma}$ and $\overline{M}$ (which is equivalent to the odd number of band crossings along the $\overline{\Gamma}$ and $\overline{M}$ points) supports the topological insulator nature of NdBi in the paramagnetic phase.

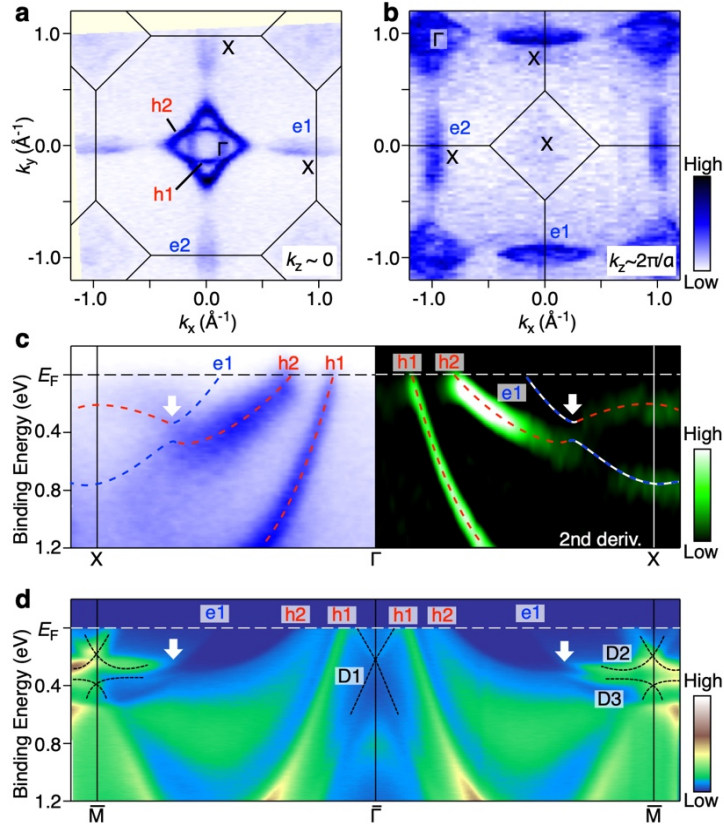

**Supplementary Fig. 1 | Bulk band structure and band inversion of NdBi. a**, **b** ARPES-intensity mapping at $E_F$ as a function of in-plane wave vector ($k_x$ and $k_y$) measured at $T$ = 40 K at $k_z \sim 2\pi/a$ ($h\nu$ = 601 eV) and $\sim$ 0 ($h\nu$ = 515 eV), respectively. **c** ARPES intensity (left) and corresponding second derivative intensity (right) plotted as a function of $k_x$ and binding energy ($E_B$) along the ΓX cut of bulk BZ measured with SX photons of $h\nu$ = 515 eV. Red and blue dashed curves are a guide for the eyes to trace the Bi-6$p$ (h1 and h2) and Nd 5$d$ (e1) bands, respectively. **d** ARPES intensity along the $\overline{\Gamma}\overline{M}$ cut of surface BZ measured at $T$ = 35 K with VUV photons of $h\nu$ = 60 eV. Black dashed curves are a guide for the eyes to trace the surface band dispersions.

**Supplementary Note 2**: **Band calculations for the PM and AF phases**

We have carried out first-principles band-structure calculations for NdBi with mBJ potential[11] and GGA potential[12] to support (i) the bulk-band inversion, (ii) the topological nature from the parity analysis, and (iii) the reproduction of D1–D3 SS in the PM phase, and (iv) the reproduction of the Dirac gap in the calculation for the AF phase. Regarding (i), we show in Fig. 1c, d of the main text a direct comparison of the bulk-band dispersion along the ΓX cut between the experiment for the PM phase ($T = 40$ K) and the calculation for the nonmagnetic phase performed with mBJ potential which is known to properly reproduce the band gap in $RX_p$ (ref. 13). Here, Nd-$4f$ electrons were treated as core states in the calculation. One can immediately recognize that the overall band structure shows a good agreement between the two. In particular, the experimental hole band topped at the binding energy ($E_B$) of ~ 1.6 eV and the inner hole band h1 are well reproduced by the calculation. In the experiment, we observe band crossing between the h2 and e1 bands and the resultant spin-orbit gap opening at the intersection. This bulk band inversion is well reproduced in the calculation. Thus, the bulk-band inversion is supported in our band calculation.

Regarding (ii), we calculated the parity eigenvalues $\xi$ for the valence bands to obtain the $Z_2$ index[14] for NdBi based on the above bulk band calculation. Since band 9 in Fig. 1d is assigned to the conduction band and bands 1–5 form fully occupied closed shells (Nd $5s$, $5p$ and Bi $6s$), it is sufficient to consider the topmost three valence bands (bands 6–8) for the parity analysis. Supplementary Table 1 shows $\xi$ at eight time-reversal invariant momenta (TRIM; $\Gamma_i$), i.e. Γ, 3X, and 4L. For each TRIM, the $\delta_i$ value was obtained by multiplying $\xi$ for bands 6–8, and the $Z_2$ topological invariants ($v_0; v_1, v_2, v_3$) were calculated as follows.

$$(-1)^{v_0} = \delta(\Gamma)\delta(X)^3\delta(L)^4 \quad = -1$$
$$(-1)^{v_1} = \delta(X)^2\delta(L)^2 \quad = 1$$
$$(-1)^{v_2} = \delta(X)^2\delta(L)^2 \quad = 1$$
$$(-1)^{v_3} = \delta(X)^2\delta(L)^2 \quad = 1$$

| $\xi, \delta_i$ \ $\Gamma_i$ | Γ | 3X | 4L | $v_0$ |
|---|---|---|---|---|
| $\xi_8(\Gamma_i)$ | + | + | – | |
| $\xi_7(\Gamma_i)$ | + | – | – | |
| $\xi_6(\Gamma_i)$ | – | – | – | |
| $\delta_i$ | – | + | – | 1 |

**Supplementary Table 1 | Parities of energy bands at 8 TRIMs.** Products of parity eigenvalues of the occupied valence-band states ($\xi$) for bands 6–8, at the time-reversal invariant momenta (TRIMs; $\Gamma_i$) of the bulk fcc BZ. $\delta_i$ is a product of $\xi$ at 8 TRIMs for all the valence bands. $v_0 = 1$ indicates the nontrivial $Z_2$ topology of NdBi in the PM phase.

Here, topological invariants $v_1$, $v_2$, and $v_3$ were calculated for the cleaved planes of (100), (010), and (001), respectively, to obtain direct correspondence with the ARPES experiments. The result shows that NdBi is a strong TI in the PM phase with ($v_0$; $v_1$, $v_2$, $v_3$) = (1; 0, 0, 0), in good agreement with the band inverted character shown in Fig. 1c.

Regarding (iii), we have carried out tight-binding calculations for 100 atomic layers in the PM phase of NdBi, and calculated the surface projected spectral weight along high symmetry lines of the surface BZ as shown in Fig. 1h of the main text (note that the Nd-4$f$ orbitals were assumed to be the core states as in the case of the bulk-band calculations because this assumption was necessary for constructing Wannier functions by fitting the calculated DFT bands to correctly simulate the surface-projected spectral weight). The result signifies the Dirac-cone SS, D1, at the $\bar{\Gamma}$ point, together with the double Dirac-cone SSs, D2 and D3, at the $\bar{M}$ point, consistent with the ARPES results for the PM phase.

Regarding (iv), we have carried out slab calculations in the AF phase with 12 atomic-layer slabs by taking into account the actual type-I AF structure (Supplementary Fig. 2a). To properly take into account the magnetic moment of Nd ions, we included the strong correlation effect of Nd-4$f$ electrons by using GGA+$U$ potential instead of mBJ potential which has a nonconvergence problem in the slab calculations as in the previous study[13]. In GGA+$U$ slab calculations, the Nd-4$f$ bands also appear in the calculation at the binding energy of 5–6 eV (out of the energy range of Supplementary Fig. 2). Since this treatment made it difficult to calculate the surface-projected spectral weight because of the difficulty in constructing the proper Wannier functions by fitting the calculated DFT bands due to the presence of 4$f$ states, we show in Supplementary Fig. 2b the original band dispersion for the 12 atomic-layer slab instead of the surface-projected weight.

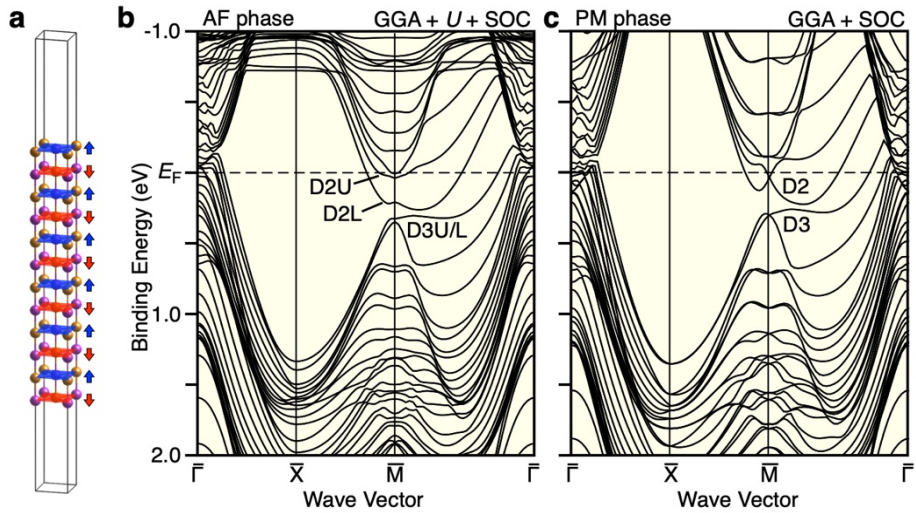

**Supplementary Fig. 2 | Calculated band structure in the AF phase. a** 12 atomic layer slab structure with type-I AF configuration adopted for the calculations to simulate the band structure for domain A. **b**, **c** Calculated band structure in the AF and PM phases along high-symmetry lines in the surface BZ. Possible D2 and D3 bands are also indicated.

Although obtaining a one-to-one correspondence in the calculated band structure with the PM state in Fig. 1h is difficult, for the top surface corresponding to domain A in the ARPES data, one can see a signature of the D2 and D3 bands near $E_F$. Intriguingly, an energy gap opens at the $\bar{M}$ point for both the D2 and D3 bands. This gap is confirmed to be of AF origin because the calculation for the PM phase obtained for the same 12 atomic-layer slab in Supplementary Fig. 2c clearly shows the gapless behavior due to the protection by the time-reversal symmetry. Also, the gap for the D2 band shown in Supplementary Fig. 2b is significantly enhanced relative to that of the D3 band. These key features are nicely reproduced in the ARPES experiment (Fig. 2h and 2i in the main text). It is noted here that the validation of the Dirac gap for the D1 band was difficult because of the overlap with the bulk bands, as shown in Fig. 1h. We have carried out slab calculations for the AF phase also for domain B (Supplementary Fig. 3a) to further validate the concept of *S*-symmetry protection. The band structure unfolded to the original BZ shown in Supplementary Fig. 3b signifies that the Dirac gap for the D2 and D3 bands is absent, distinct from domain A (Supplementary Fig. 2). This theoretically demonstrates the selectively gapped Dirac-cone state and supports the validity of *S*-symmetry protection. Such an intriguing surface-dependent Dirac gap is overall consistent with the ARPES observation shown in Fig. 3a–f, whereas the origin of the residual gap for the D2 band in domain B (Fig. 3f) remains as an open question. The deviation from the theory may invoke exotic mechanisms beyond the present DFT calculation, such as modulation of the surface magnetic structure, spin fluctuations, and strong many-body effects. We leave this issue as a challenge in the future study.

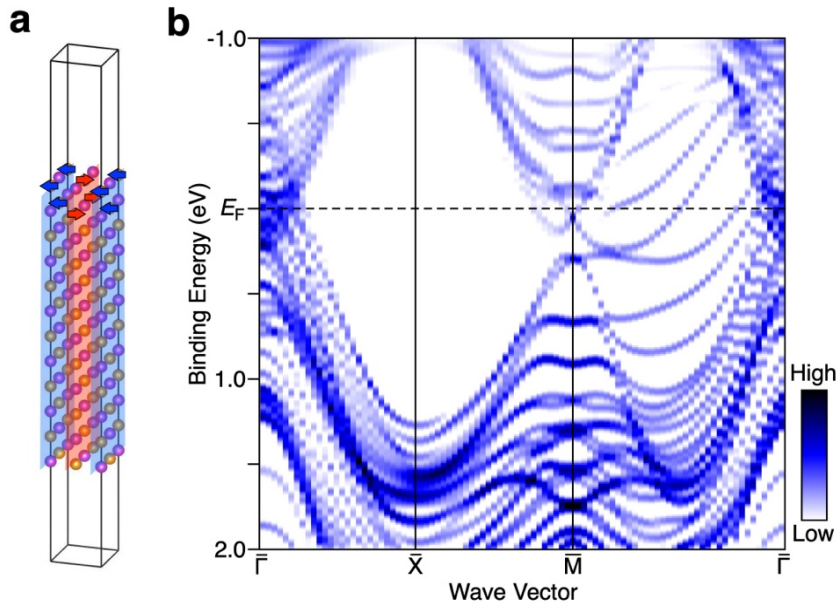

**Supplementary Fig. 3 | Calculated band structure in the AF phase for domain B. a** 12 atomic layer slab model with type-I AF configuration for domain B. √2×√2 supercell is adopted to simulate the transverse AF order. **b** Intensity of spectra corresponds to the amplitude of projection to the original cell eigenstate.

All these arguments regarding (i) the bulk-band inversion commonly identified in the experiment and calculation, (ii) the strong TI [$(v_0; v_1, v_2, v_3) = (1; 0, 0, 0)$] nature in the PM phase from the parity analysis, (iii) the reproduction of D1–D3 SS by the slab calculations, and (iv) the Dirac-gap opening for the calculated D2 and D3 SS in the AF state for domain A, support and strengthen our main claim on the AF TI nature of NdBi.

**Supplementary Note 3**: $k_z$-independent energy dispersion of the Dirac-cone SS

We have carried out $hv$-dependent ARPES measurements to experimentally demonstrate the surface nature of the Dirac-cone state by measuring the $k_z$ dispersion with several photon energies ($hv$'s). Since the intensity of D1–D3 bands was found to be strongly suppressed for the bulk-sensitive soft X-ray (SX) photons (see Supplementary Fig. 1), we measured the detailed $hv$ dependence with the surface-sensitive vacuum ultraviolet (VUV) photons at $hv$ = 44, 50, 56.5, 60, 63, 70 eV, besides $hv$ = 75 eV (Fig. 2c of the main text). This $hv$ range fully covers the ΓX length of the bulk Brillouin zone ($k_z$ = 0 to $2\pi/a$) with a reasonably small $k_z$ step. We found from the ARPES intensity obtained in the PM phase at $T$ = 40 K shown in Supplementary Fig. 4 that the energy position of the D1–D3 bands estimated from the ARPES data at $hv$ = 75 eV (same as dashed curves in Fig. 2c and 2h of the main text) overlaps with the intensity of the D1–

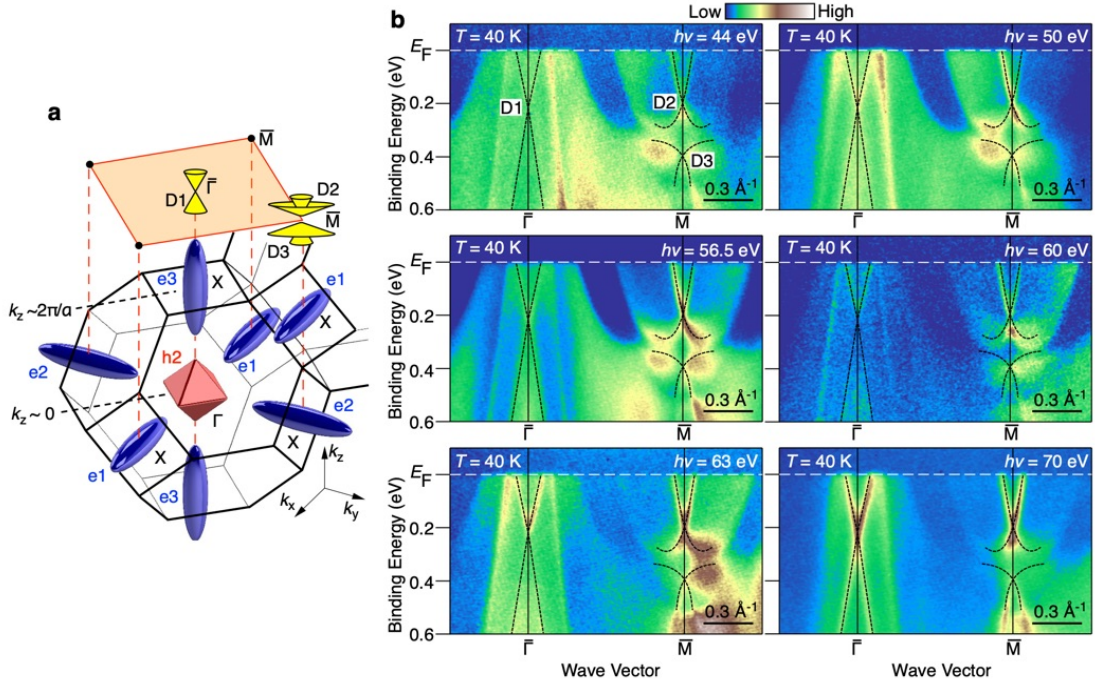

**Supplementary Fig. 4 | $k_z$-independent energy dispersion of the Dirac-cone SS in NdBi. a** Schematic FS and bulk fcc BZ of NdBi, together with the surface BZ projected onto the (001) plane (orange rectangle) and Dirac-cone SS at the $\bar{\Gamma}$ and $\bar{M}$ points (same as Fig. 2c in the main text). **b** $hv$-dependence of the ARPES intensity along the $\bar{\Gamma}\bar{M}$ cut of surface BZ, measured in the PM phase ($T$ = 40 K) at $hv$ = 44–70 eV. Dashed curves are a guide for the eyes that traces the experimental band dispersion of the D1–D3 bands obtained at $hv$ = 75 eV (same as Fig. 2c, h in the main text).

**Supplementary Note 4**: **AF origin of the Dirac gap for the D1 band**

We have carried out detailed temperature-dependent ARPES measurements for the D1 band to clarify whether the energy splitting in the AF phase is indeed associated with the AF transition. As shown in Supplementary Fig. 5, one can recognize a clear energy splitting of the upper Dirac-cone (D1U) and lower Dirac-cone (D1L) bands at $T = 15$ K (Supplementary Fig. 5a1). On increasing temperature, these bands gradually broaden and approach each other. The splitting appears to persist at least up to $T = 21$ K (Supplementary Fig. 5a7) and becomes invisible at 24 K (= $T_N$) and 27 K. As shown by the ARPES intensity at the $\bar{\Gamma}$ point plotted against temperature in Supplementary Fig. 5b, the D1 band starts to split into the D1L and D1U bands just at $T_N$. These results strongly support the AF origin of the Dirac gap.

It is noted here that the energy position of the DP in the PM phase (0.21 eV) in Fig 2c of the main text (as represented by the ARPES data in Supplementary Fig. 5a11) estimated by the linear extrapolation of MDC peak positions slightly deviates from the peak position of the EDC at the $\bar{\Gamma}$ point in Fig. 2e (0.18 eV). We think that this difference is associated with the local deviation of the Dirac-band dispersion from the linear behavior around the DP. The upper Dirac-cone band is rounded around the Dirac point and connected to a highly dispersive lower Dirac-cone band with weaker intensity at the $\bar{\Gamma}$ point. Indeed, such behavior was recognized in previous ARPES and DFT-calculation studies in other rare-earth monopnictides[2–5]. Another mechanism to cause a deviation of the DP and EDC-peak energies may be a complication of the overall spectral lineshape (in particular EDCs) due to the strong surface-bulk interaction for the D1 band as supported by our first-principles band calculations.

Here we briefly discuss the reason why the observed gap (125 meV) in the AF state is so large. The gap size in NdBi is indeed much larger than those in other AF TI candidates such as $MnBi_2Te_4$ (MBT) which shows the AF-induced gap of at most 85 meV[15–17] (note that the reported gap size significantly varies depending on the group). While the exact reason for the observed large gap is unclear at the moment, we speculate that the Dirac gap size is linked to the density and magnetic moment of magnetic ions (Nd/Mn). Specifically, the magnetic moment for $Nd^{3+}$ ion (3.1 $\mu_B$) is ~70 % of that of $Mn^{2+}$ ion (4.6 $\mu_B$), while the magnetic ion density in NdBi (Nd:Bi = 1:1) is 350 % of that of MBT (Mn:$Bi_2Te_4$ = 1:6). In total, the effective exchange field for the surface Dirac electrons is expected to be much larger in NdBi than in MBT, likely producing a larger gap in NdBi. However, the exchange constants estimated from the neutron diffraction experiments are $J_1 = -0.008$ meV and $J_2 = 0.016$ meV for the nearest neighbor and next nearest neighbor, respectively[18], far smaller than the observed Dirac gap. Thus, although a qualitative argument on the larger Dirac gap may be possible, it is hard to quantitatively explain the magnitude of the Dirac gap in NdBi.

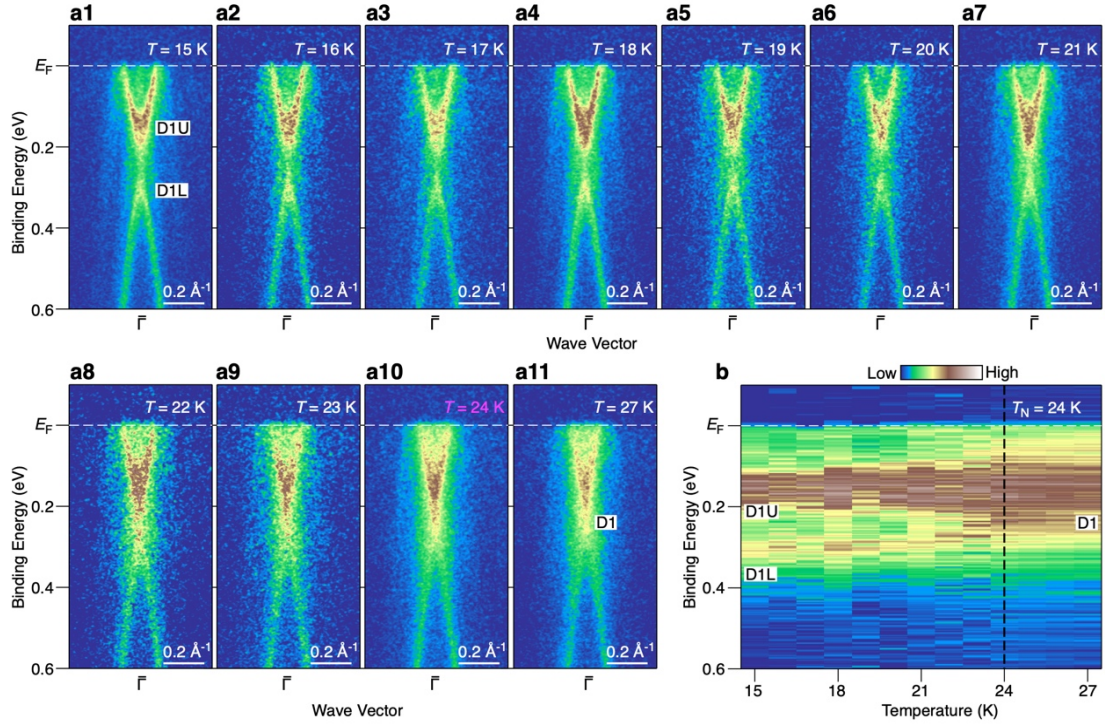

**Supplementary Fig. 5 | Temperature dependence of the D1-band dispersion across $T_N$. a1–a11** ARPES intensity along the $\overline{\Gamma M}$ cut for the D1 band measured at representative temperatures ($T$ = 15–27 K) across the Néel temperature $T_N$ (= 24 K) measured at $h\nu$ = 75 eV. X-shaped gapless Dirac-cone band (D1) in the PM phase turns into the gapped Dirac-cone band (D1U and D1L) in the AF phase. **b** Temperature dependence of the ARPES intensity at the $\overline{\Gamma}$ point which signifies the energy splitting of the D1 band in the AF phase.

**Supplementary Note 5**: **Identification of AF domains by polarizing microscopy**

We have characterized the AF domain structure of NdBi to substantiate our main claim regarding the AF-domain-dependent Dirac-cone feature by using a low-temperature UHV polarizing microscope system as shown in Supplementary Fig. 6a, b. This system is based on the birefringence of the sample associated with the coupling of optical properties and electronic states, and detects the change in the light polarization between incoming and outgoing reflected photons (Supplementary Fig. 6c), as already applied to other rare-earth monopnictides to detect AF domains[19]. In this system, white light emitted from a halogen lamp is horizontally polarized, and irradiated on the sample. Then, the reflected light passes the vertical polarizer and is detected by a CCD camera (crossed Nicols configuration). Depending on the direction of the Nd-4$f$ magnetic moment with respect to the polarization vector of the incident light, three types of AF domains can be distinguished in this system[16]. Supplementary Fig. 6g shows a microscope image subtracted between the PM (Supplementary Fig. 6e) and the AF phases (Supplementary Fig. 6f). One can see domain A with the out-of-plane magnetic moment as a white colored region (i.e. no birefringence) and domain B with the in-plane magnetic moment as a red/blue colored region. Importantly, we found that the spot size of the microbeam shown by the black circle is much smaller than the typical AF domain size. This result supports that the spatially dependent band structure observed by ARPES is related to the difference in the AF domain.

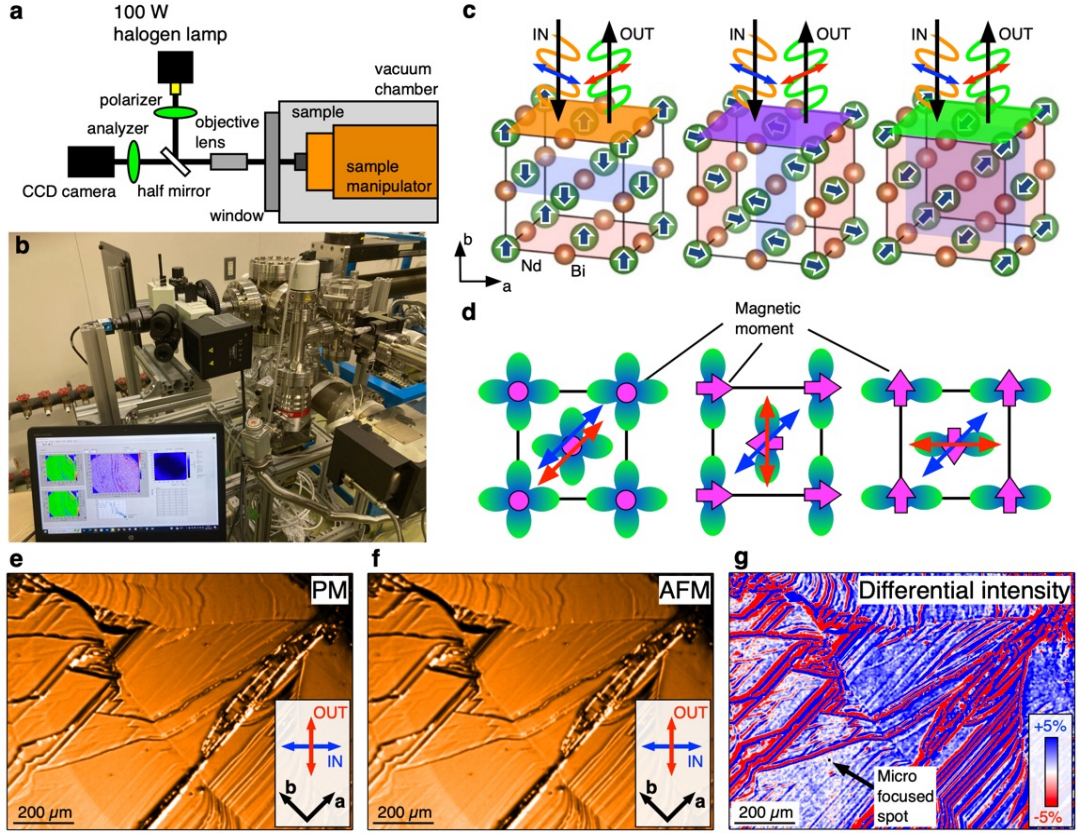

**Supplementary Fig. 6 | Polarizing microscopy measurements in the AF phase. a** Schematic of a constructed polarizing microscopy system. **b** Photograph of the polarizing microscope system. **c** Three types of AF domains in NdBi. **d** Schematics of light polarizations for outgoing photons (red arrows) that have different rotation angles with respect to that of incident photons (blue arrows) depending on the type of AF domains. **e, f** Microscope image obtained at $T = 30$ K (PM phase) and 8 K (AF phase), respectively. **g** Polarizing microscope image obtained by subtracting the images between $T = 30$ K and 8 K.

**Supplementary Note 6**: **Observation of anisotropic electronic states for domain B**

To clarify the symmetry of band dispersion and Fermi surface in the AF phase for domain B, we show in Supplementary Fig. 7a, b the Fermi-surface mapping at $T = 5$ K around the $\bar{\Gamma}$ and $\bar{M}$ points, respectively. Corresponding ARPES intensity plots as a function of wave vector ($k_x$ or $k_y$) and binding energy ($E_B$) obtained along representative **k** cuts (cuts 1–4) are also shown in Supplementary Fig. 7c–f. One can see from the experimental band dispersions along $k_x$ and $k_y$ cuts (cuts 1 and 2) across the $\bar{\Gamma}$ point (Supplementary Fig. 7c, d) that shallow bands in the vicinity of the Fermi level forming the S2 pocket appear along the $k_x$ cut (cut 1), but not along the $k_y$ cut (cut 2). Similar inequivalence between the vertical and horizontal **k** cuts is also observed around the $\bar{M}$ point, as shown in Supplementary Fig. 7e, f. This demonstrates the $C_2$ symmetric nature of the band structure and Fermi surface for domain B, supportive of the AF-ordering vector parallel to the in-plane $k_x$ direction, in sharp contrast to the $C_4$ symmetric Fermi surface for domain A (Fig. 4a, b in the main text).

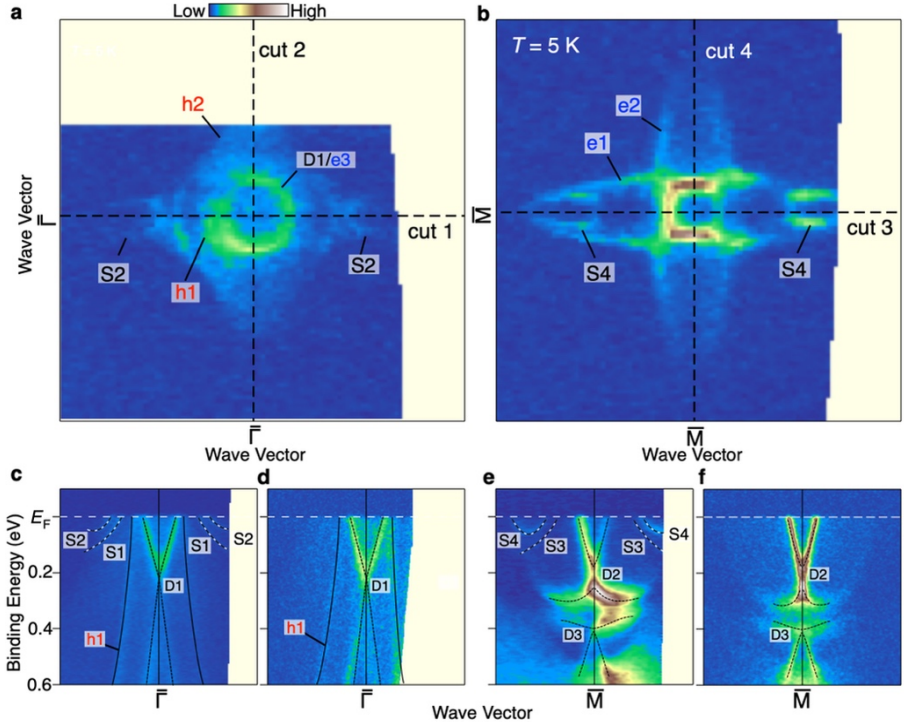

**Supplementary Fig. 7 | $C_2$ symmetric Fermi surface for domain B. a**, **b** Fermi-surface mapping around the $\bar{\Gamma}$ and $\bar{M}$ points, respectively, for domain B (same as Fig. 4c, d, respectively). **c**–**f** ARPES intensity in the AF state ($T$ = 5 K) for domain B, measured along four representative **k** cuts (cuts 1–4) shown in **a** and **b**.

**Supplementary Note 7**: **Intrinsic nature of the $C_2$ symmetric electronic states for domain B**

We have carried out ARPES measurements with different matrix-element conditions and excluded the matrix-element effect from the origin of the observed $C_2$ symmetric electronic structure for domain B. Supplementary Fig. 8a, b compare the Fermi surface mapping for domain B obtained with different light polarizations (linear vertical vs circular) and photoelectron emission angles (2$^{nd}$ BZ vs 1$^{st}$ BZ) as shown in the insets. Although the ARPES intensity is strongly modulated by the matrix-element effect, one can still recognize small S2 pockets along the $k_x$ cut but not along the $k_y$ cut. Such a $C_2$ symmetric feature is better visualized by the comparison of the ARPES intensity along the $k_x$ and $k_y$ cuts in which the pocket is only seen along the $k_x$ cut irrespective of the difference in the matrix-element effects (Supplementary Fig. 8c–f). This result suggests that the observed $C_2$ symmetric electronic structure is an intrinsic feature of domain B.

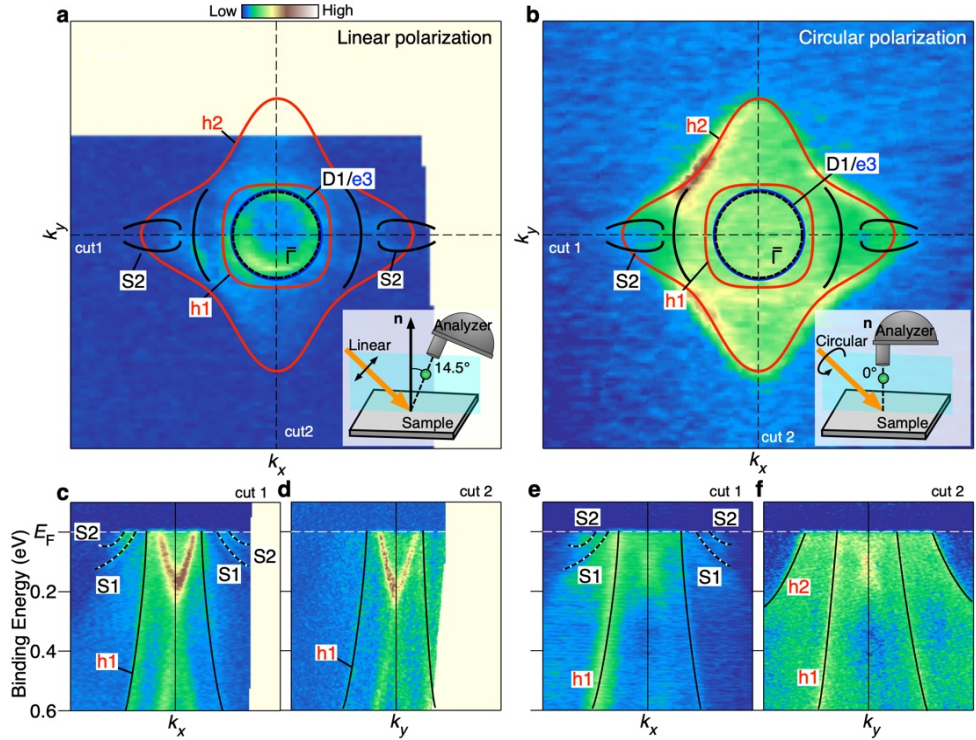

**Supplementary Fig. 8 | Light-polarization- and geometry-independent $C_2$ symmetric nature of Fermi surface for domain B. a** ARPES-intensity mapping at $E_F$ as a function of $k_x$ and $k_y$ in the AF phase ($T$ = 5 K) measured with linear vertically polarized light ($hv$ = 60 eV) with the measurement geometry shown in the inset which covers the $\bar{\Gamma}$ point of second surface BZ. Black solid curves are a guide for the eyes to trace the experimental Fermi surface. **b** Same as **a** but measured with circularly polarized light with the measurement geometry (inset) which covers the $\bar{\Gamma}$ point of first surface BZ. **c, d** ARPES intensity plotted as a function of wave vector ($k_x$ for cut 1 and $k_y$ for cut 2) and $E_B$. **e, f** Same as **c, d** but for the measurement geometry of **b**.